\documentclass{aa}
\usepackage{graphicx}
\usepackage{subfigure}
\usepackage{natbib}
\usepackage{txfonts}
\bibpunct{(}{)}{;}{a}{}{,}

\begin{document}

\title{Identification of 13 DB~+~dM and 2 DC~+~dM binaries from the
Sloan Digital Sky Survey}
\titlerunning{Identification of 15 non-DA~+~dM binaries from the SDSS.}

\author{E.J.M. van den Besselaar\inst{1} \and G.H.A. Roelofs\inst{1} 
  \and G.A. Nelemans\inst{1} \and T. Augusteijn\inst{2} 
  \and P.J. Groot\inst{1} }

\offprints{E.J.M. van den Besselaar, \\
\email{besselaar@astro.ru.nl}}

\institute{Department of Astrophysics, Radboud University Nijmegen,
P.O Box 9010, 6500 GL Nijmegen, The Netherlands \\
\email{gijsroel@astro.ru.nl; nelemans@astro.ru.nl; pgroot@astro.ru.nl}
\and Nordic Optical Telescope, Apartado 474, Santa Cruz de La Palma, 
  Spain \\ \email{tau@not.iac.es}}

\date{Received \ldots / Accepted \ldots}

\abstract{We present the identification of 13 DB~+~dM binaries and 2
DC~+~dM binaries from the Sloan Digital Sky Survey (SDSS). Before the
SDSS only 2 DB~+~dM binaries and 1 DC~+~dM binary were known. At least
three, possibly 8, of the new DB~+~dM binaries seem to have white
dwarf temperatures well above 30\,000~K which would place them in the
so called DB-gap.  Finding these DB white dwarfs in binaries may
suggest that they have formed through a different evolutionary channel
than the ones in which DA white dwarfs transform into DB white dwarfs
due to convection in the upper layers.
\keywords{binaries: spectroscopic -- white dwarfs -- Stars:
late-type -- Stars: evolution} }

\maketitle

\vspace{-0.2cm}
\section{Introduction}
Our understanding of binary evolution is still severely lacking on a
number of points, but most importantly on the physics of
common-envelope (CE) e\-vo\-lu\-tion \citep{paczynski}. When binaries
are close enough that during their evolution Roche lobe overflow can
commence from a convective giant to a less massive main-sequence
companion star, very quickly the system will evolve into a state where
the envelope of the giant encompasses both objects.  After the
envelope has been expelled a close binary may result that will evolve
into a white dwarf -- main-sequence binary. Due to its short-lived
phase and the intrinsic three-dimensional hydrodynamic nature of the
CE phase the physics of this process is not well understood
\citep[e.g.][]{cetimescale}.

One way of improving our knowledge of the CE phase is to determine the
space densities and population characteristics of all CE
products. Some of these products are white dwarfs (WD) with a low-mass
main-sequence (dM) companion. WD~+~dM binaries (or pre-Cataclysmic
Variables), containing a hydrogen-rich white dwarf (DA), are a fairly
well known group of objects of which new members are readily found,
among others by the efforts of the Sloan Digital Sky Survey
\citep[SDSS,][]{york, raymond}. On the contrary, WD~+~dM binaries
containing a helium-rich white dwarf (DB) showing composite spectra
are extremely rare, and from literature before the SDSS only two such
objects are known, namely \object{GD 325} \citep{gd325} and
\object{CBS 47} \citep{cbs47}.  Identification of objects in these
classes may not only help in understanding the CE physics and the
evolution of Cataclysmic Variables (CVs), but also in the formation of
DBs in the first place.

For single WDs in the SDSS it is found that 9-15\% of all systems have
helium dominated atmospheres \citep{kleinman,wdsloan}. In contrast,
the fraction of WDs in known binaries that are DBs is $\ll1\%$.  DC
white dwarfs \citep[with absorption lines less than 5\% of the
continuum;][]{wdatlas} + dM binaries are an even rarer class with
only one member known \citep[\object{EG 388}; ][]{eg388}.  We here
report on the discovery of 13 new DB~+~dM binaries and 2 DC~+~dM
binaries from the SDSS.

\vspace{-0.2cm}
\begin{table*}[!th]
  \centering
  \caption{ The characteristics of our thirteen DB~+~dM and two
    DC~+~dM binaries are given in this table. The uncertainties are
    about one subtype for the dM star and about 4000 Kelvin in WD
    temperature for a fixed dM spectral type. The temperatures are
    shown as given in \citet{kleinman} for the best fit template
    WD. $^a$ See Sect.~\ref{sec:notes}.  $^b$ These objects are
    DC~+~dM binaries.}
  \begin{tabular}{l r r r r r r r r r}
    \hline
    \hline
    Name &  \multicolumn{1}{c}{$g$} & $u-g$ & $g-r$ & $r-i$ & $i-z$ & T$_{\rm WD}$ & dM & D (pc) & $\chi^2$ \\
    \hline 
    \object{SDSS J075235.79+401339.0}       & 20.01 & 0.05 & 0.01 & 0.58 & 0.39 & 30\,252 & M3V & 1544 &  1.64 \\
    \object{SDSS J080636.85+251912.1}     & 19.61 & 0.07 & 0.21 & 0.79 & 0.60 & 24\,266 & M3V & 1045 &  1.68 \\
    \object{SDSS J093645.14+420625.6}$^a$  & 20.37 & $-$0.09 & $-$0.14 & 0.42 & 0.07 & 15\,919 & M5V & 860 & 2.01\\
    \object{SDSS J100636.39+563346.8}     & 19.42 & 0.10 & 0.28 & 0.72 & 0.49 & 14\,575 & M4V &  531 &  3.70 \\ 
    \object{SDSS J102131.55+511622.9}     & 18.27 & $-$0.25 & $-$0.37 & 0.14 & 0.35 & 30\,252 & M4V &  700 &  1.56 \\
    \object{SDSS J113609.59+484318.9}$^a$ & 16.80 & $-$0.38 & $-$0.54 & $-$0.16 & 0.14 & 38\,211 & M6V &  354 &  4.00 \\
    \object{SDSS J134135.23+612128.7}     & 19.14 & $-$0.10 & 0.12 & 0.50 & 0.48 & 30\,694 & M3V & 1054 &  2.20 \\
    \object{SDSS J143222.06+611231.1} & 18.53 & 0.02 & $-$0.08 & 0.49 & 0.60 & 36\,815 & M3V &  879 &  3.56 \\
    \object{SDSS J144258.47+001031.5}$^a$     & 18.35 & 0.04 & 0.08 & 0.80 & 0.72 & 30\,694 & M3V &  674 & 11.22 \\
    \object{SDSS J150118.40+042232.3}$^a$     & 19.57 & 0.02 & $-$0.01 & 0.49 & 0.48 & 26\,020 & M3V & 1220 &  2.42 \\
    \object{SDSS J162329.50+355427.2}     & 18.77 & 0.02 & 0.22 & 0.65 & 0.53 & 24\,266 & M3V &  695 &  3.46 \\
    \object{SDSS J220313.29+113236.0}     & 19.34 & $-$0.08 & $-$0.10 & 0.22 & 0.45 & 30\,694 & M4V & 1085 &  2.04 \\
    \object{SDSS J232438.31$-$093106.5}    & 18.64 & $-$0.08 & $-$0.16 & 0.41 & 0.66 & 36\,815 & M3V &  911 &  3.45 \\
\hline
    \object{SDSS J074425.42+353040.8}$^{a,b}$ & 18.91 & 0.23 & 0.03 & 0.70 & 0.68 & 15\,700 & M4V &  481 &   7.14 \\ 
    \object{SDSS J113457.72+655408.7}$^{a,b}$ & 18.15 & 0.10 & 0.02 & 0.70 & 0.75 & 12\,500 & M4V &  278 &   9.59 \\
    \hline
  \end{tabular}
  \label{tab:wdrd}
\end{table*}

\vspace{-0.2cm}
\section{Observations}
\label{sec:obs}
\subsection*{Colour selection of WD~+~dM binaries}
\vspace{-0.2cm}
\begin{figure}
  \centering
  \resizebox{\hsize}{!}{\includegraphics[angle=270]{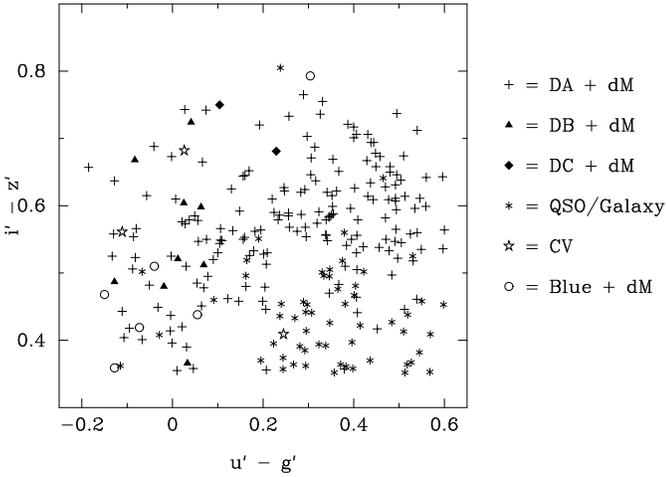}}
    \caption{A $u-g$ vs. $i-z$ diagram of our 260 objects together
    with their classification based on visual inspection of all the
    spectra satisfying our selection criteria.}
    \label{fig:colour}
\end{figure}

We have used the spectroscopic data from the Third Data Release of the
SDSS (DR3) to select WD + dM candidates. Because of their composite
spectra these binaries stand apart from normal main-sequence stars in
any colour -- colour diagram \citep[see e.g. Fig.~1 of][]{raymond}.
We have used the following colour criteria for selecting candidates:
$-0.2<u^{\prime} - g^{\prime}<0.6$, $-0.2<g^{\prime} -
r^{\prime}<0.3$, $0.35<r^{\prime} - i^{\prime}<0.85$ and
$0.35<i^{\prime} - z^{\prime}<0.85$, where the magnitudes are the
psf-magnitudes taken from the spectrophotometric table of DR3.  These
criteria were derived from colour -- colour diagrams of an initial set
of WD + dM binaries that were selected from the spectroscopic
database.

Using these colour criteria, we selected 260 objects. We classified
them on the basis of visual inspection of their spectra. Of these
objects 58 were classified as quasars or galaxies. A further 6 were
too noisy to be classified, leaving 196 dM binaries. These are further
classified as 176 DA~+~dM binaries, 9 DB~+~dM binaries, 2 DC~+~dM
binaries, 3 Cataclysmic Variables and 6 dM binaries with a blue
component too noisy to classify. A $u^{\prime}-g^{\prime}$
vs. $i^{\prime}-z^{\prime}$ diagram is shown in Fig.~\ref{fig:colour}.

\subsection*{Equivalent width selection of DB~+~dM binaries}
A second, independent, search for DB + dM binaries was performed by
looking for He absorption lines in all DR3 spectra. Apart from single
DBs this should select all DB + dM binaries in which the DB dominates
the blue part of the spectrum.  In practice, this means selecting
binaries in which the DB has a T$_{\rm eff}\geq$20\,000 K with a
companion of spectral type M1 or later, and binaries in which the DB
has a T$_{\rm eff}\geq$14\,000 K with a companion of spectral type M3
or later.  For all spectra we calculated the equivalent width (EW) of
the \ion{He}{I} 4026, 4471, 4921 and 5876~\AA~lines. On the basis of
the distributions of EWs in a sample of 160 bright DBs and requiring
that all four lines are present, DB candidates were selected such that
less than an estimated 5\% of all DBs fall outside the selection due
to weak absorption lines.

All resulting candidates from DR3 were visually inspected for DB + dM
signatures. The 9 DB + dM binaries from our colour selection were
re-discovered and 4 additional systems (outside our colour selection)
were identified, increasing the total number of these rare systems
found in the SDSS to 13.

\vspace{-0.2cm}
\begin{figure*}
  \centering 
  \includegraphics[angle=270, width=17cm]{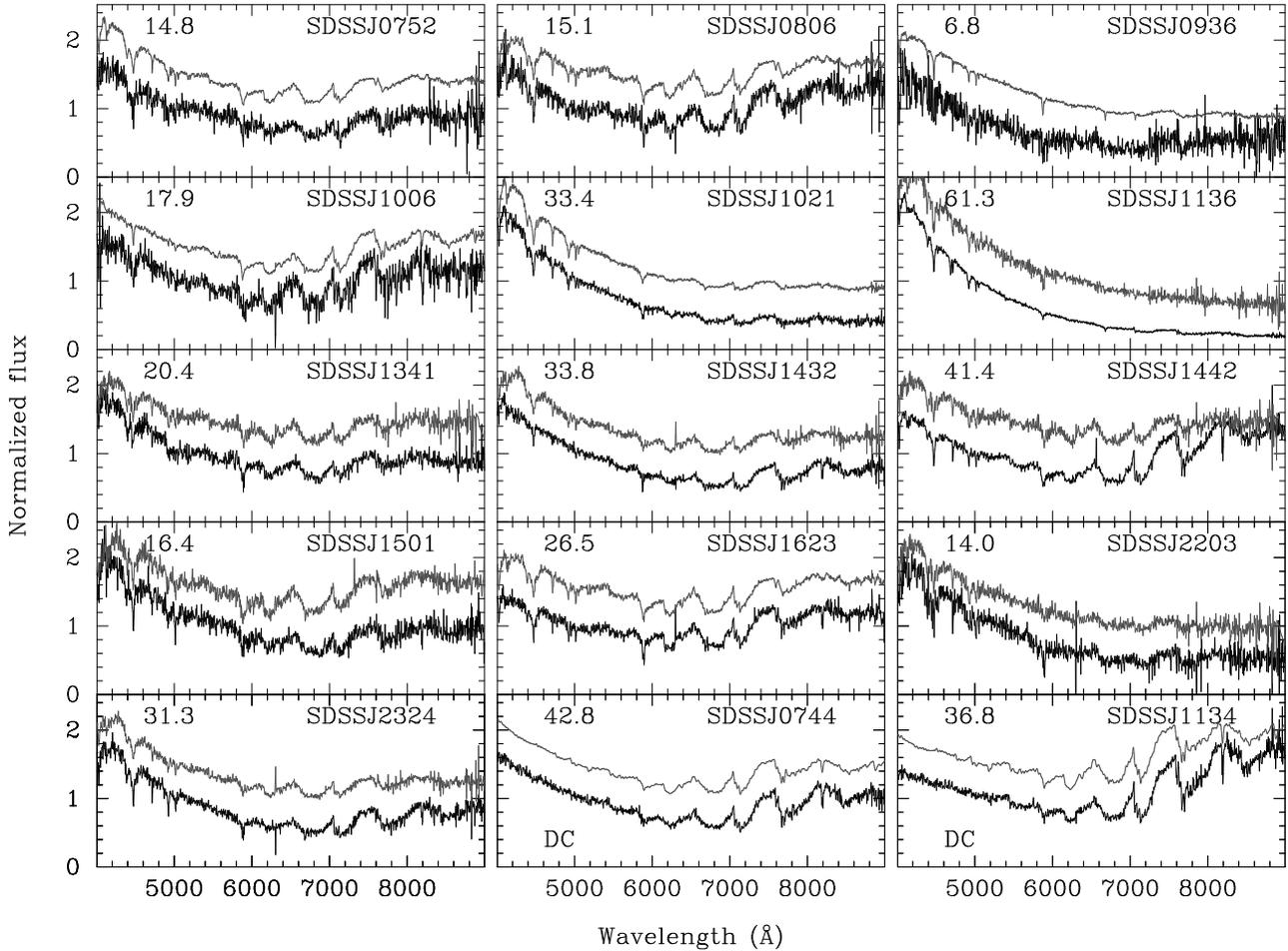}
  \caption{The 13 DB + dM and 2 DC + dM (bottom row) systems
    (Table~\ref{tab:wdrd}). The object spectra are shown as the lower
    spectra, the best template has been shifted upwards for
    clarity. The number in the upper left corner of each plot is the
    S/N in a 5~\AA~bin at 4600~\AA.}
  \label{fig:spectra}
\end{figure*}

\vspace{-0.2cm}
\section{Results}
\label{sec:results} 
We fitted the DB~+~dM spectra with a combined template DB and template
dM spectrum using the $\chi^2$ minimalization method.  For the fit we
have scaled the flux in the template spectra to a distance of 10 pc
before combining them. The 17 template DB spectra and their
corresponding temperatures, between 13390 and 38211~K, were taken from
\citet{kleinman}. To derive the temperature of the DC white dwarfs, we
have fitted blackbody spectra with a tempe\-rature range of 8\,000 --
40\,000 K. The 7 dM spectra ranging from type M0V to M6V were taken
from \citet{pickles}.

The best fit Kleinman-temperatures (or blackbody temperatures in case
of a DC + dM), spectral types, and $\chi^2$ for the 15 non-DA~+~dM
binaries are given in Table~\ref{tab:wdrd}. In most cases, the
uncertainty in spectral type of the dM is 1 subtype or less. For a
fixed dM the uncertainty on the temperature of the WD is estimated to
be about 4000~K. However, there is a correlation between the WD
temperature and the dM spectral type. For a fit with a dM one subtype
later, the best template WD may have a 8\,000 -- 10\,000 K lower
temperature with slightly increased $\chi^2$. The spectra are shown in
Fig.~\ref{fig:spectra} (hereafter we will abbreviate names to the form
SDSSJhhmm). The object spectra are the lower spectra, while the
template spectra are the upper ones.

In this paper we give all the parameters as derived assuming a WD with
a radius of 0.0123~R$_{\sun}$, identical to \citet{raymond}. The
mass-radius relation from Eggleton as quoted by \citet{verbunt} gives
a corresponding mass of 0.6~M$_{\sun}$ for this radius.  We used
several WD radii between 0.007 and 0.02~R$_{\sun}$ (a mass between
1.08 and 0.21 M$_{\sun}$, respectively) in all our fits. For the
DB~+~dM binaries the WD temperature, dM spectral type and overall
$\chi^2$ change only very slightly with different radii. However, for
the DC~+~dM binaries the fit improves for a WD with a radius of
0.008~R$_{\sun}$ (a mass of about 1~M$_{\sun}$).

\smallskip
\noindent
\label{sec:notes}
\textbf{SDSSJ0936\\} 
Due to the low signal-to-noise the blue component in SDSSJ0936 is
tentatively classified as a DB, but a better spectrum is needed to
confirm this.

\smallskip
\noindent
\label{sec:sdss1136}
\textbf{SDSSJ1136\\} 
SDSSJ1136 shows \ion{He}{ii} absorption at 4686~\AA~ and possibly at
5411~\AA~indicating a high WD temperature.  Close inspection shows
clear features of a cool companion to the WD. Our best fit to this
spectrum is with the combination of a WD with a temperature of 38\,211
K and an M6 companion which are our hottest template WD and coolest
template dM spectra. The $\chi^2$ is still rather high (see
Table~\ref{tab:wdrd}), suggesting at least a WD with a temperature
$>$~38\,000~K and possibly a companion with a spectral type later than
M6V.

\smallskip
\noindent
\label{sec:sdss1442}
\textbf{SDSSJ1442\\} 
The best fit to this object is still rather poor. If we decrease the
radius of the WD in the fit to 0.009~R$_{\sun}$, the fit result is
significantly better. \citet{raymond} have included this object in
their paper but they only mention that it has He absorption. In their
Table 1 they have given a temperature of 32\,000 K, but this is based
on a hydrogen WD model. In Table~1 of \citet{wdsloan} SDSSJ1442 is
classified as a DB3.5 + M binary but it is not discussed further in
their paper either.  Furthermore, \citet{raymond} have taken 2
follow-up spectra of this object, because it shows signs of
chromospheric activity through strong H$\alpha$ emission, and find a
minimum radial velocity variation of 150~km~s$^{-1}$. This suggests
that SDSSJ1442 is a close binary.

\smallskip
\noindent
\label{sec:sdss1501}
\textbf{SDSSJ1501\\} 
The object SDSSJ1501 is mentioned in the table of \citet{raymond} but
is not mentioned elsewhere in that paper. They have fitted a hydrogen
WD model to this spectrum, which might (partly) explain the difference
in temperature and dM spectral type between our results and
theirs. SDSSJ1501 clearly shows He absorption and no hydrogen
absorption and therefore should be classified as a DB~+~dM binary.

\smallskip
\noindent
\label{sec:sdss0744}
\textbf{SDSSJ0744 \& SDSSJ1134\\} 
The blue parts of the spectra of SDSSJ0744 and SDSSJ1134 show no
absorption lines from the WD stronger than 5\% of the continuum and
are therefore classified as DC~+~dM binaries.  The object SDSSJ1134 is
mentioned in the table of \citet{raymond} as well and was also fitted
with a hydrogen WD model. If we use a smaller radius which corresponds
to a WD mass of about 1~M$_{\sun}$ the fits improve to
$\chi^2\approx4$. This results in later type M dwarf secondaries.

\vspace{-0.2cm}
\section{Discussion}
\label{sec:discussion}
We have identified 13 DB~+~dM binaries and 2 DC~+~dM binaries in the
SDSS DR3 and we derived the WD temperature, dM spectral type and
distance of these binaries.

Almost no single DBs have been found with temperatures between 30\,000
and 45\,000~K, the so called DB-gap \citep{liebert}. \citet{kleinman}
have found some DBs with temperatures above 30\,000~K in the SDSS, but
only 11 out of 171 single DBs. Although better spectra and modeling
with WD atmospheric models is needed, 3 of our 13 DB~+~dM binaries
appear to have temperatures well above this limit and another 5 have
temperatures around 30\,000~K.  The fraction of DBs in the DB-gap for
these binaries is therefore very high compared to the total DB
population. From the fact that we do find cool DBs as well, it is
clear that this is not due to our selection method. As mentioned
before, if the fitted dM spectral type is 1 subtype too early, the
temperature may drop with 8\,000 -- 10\,000 K. Even in this worst-case
scenario, still at least 3 binaries will have temperatures close to
the DB-gap. SDSSJ1136 shows \ion{He}{ii} absorption which suggest a
temperature of above 30\,000 K, placing it in the DB-gap. Follow-up
observations of all 13 DB~+~dM systems is needed to obtain high S/N
spectra. Then we can model these spectra with WD atmospheric models to
be certain which part of this sample is in the DB-gap.

It is thought that during the cooling sequence DAs can transform into
DBs due to convective mixing in the H layer for temperatures below
30\,000 K \citep[e.g.][]{kalirai}. WDs in close binaries might accrete
some matter from their companions, making their H layers too thick for
the transition to DBs. However, the fact that we do find DBs in
(close) binaries may imply that they are formed in a different way so
that they have lost their H (almost) completely. This could also
explain their existence in the DB-gap and would suggest the DB/DA
ratio in binaries might be different from that in single WDs.

The DB/DA fraction for single WDs in the SDSS is 9-15\%
\citep{kleinman,wdsloan}. To derive the fraction of DB/DA in binary
systems we can not use the EW selection. The He lines in DBs are very
useful in an EW selection, but this is not possible for H lines due to
the dominance of main-sequence stars.  Therefore we use the sample
based on the colour selection to derive the DB/DA ratio in binary
systems, yielding 5\%. Due to the colour selection this could be
biased, but the nine DB~+~dM binaries that we selected from the colour
selection show a similar spread in temperature as the other DB
binaries in the sample. This indicates that our colour selection does
not specifically bias towards binaries with certain WD temperatures,
indicating that the ratio is different in binaries than in single
stars, though a more detailed study is necessary.

\citet{raymond} estimate that about 5\% of the WD + dM are close
binaries with short orbital periods.  They have taken 2 follow-up
spectra of SDSSJ1442 from which a minimum radial velocity variation of
150~km s$^{-1}$ can be derived.  From this value it can be assumed
that SDSSJ1442 is a close binary. H${\alpha}$ emission in close
binaries can be enhanced due to faster rotation of the secondary or
due to strong heating of the secondary by the WD, so H${\alpha}$
emission can be an indication of a close binary
\citep[e.g.][]{raymond}. In our sample there are 3 systems (SDSSJ1341,
SDSSJ1432 and SDSSJ1442) that show H${\alpha}$ emission.  Follow-up
observations of all the sources in our sample is needed to determine
if they are close binaries, and to compare their periods to those of
DA~+~dM binaries to investigate the formation channel.

\begin{acknowledgements}
EvdB, GR and PJG are supported by NWO-VIDI grant 639.042.201 to
P.J. Groot. GN is supported by NWO-VENI grant 639.041.405 to
G. Nelemans.

Funding for the creation and distribution of the SDSS Archive has been
provided by the Alfred P. Sloan Foundation, the Participating
Institutions, the National Aeronautics and Space Administration, the
National Science Foundation, the U.S. Department of Energy, the
Japanese Monbukagakusho, and the Max Planck Society. The SDSS Web site
is http://www.sdss.org/.  The SDSS is managed by the Astrophysical
Research Consortium (ARC) for the Participating Institutions. The
Participating Institutions are The University of Chicago, Fermilab,
the Institute for Advanced Study, the Japan Participation Group, The
Johns Hopkins University, the Korean Scientist Group, Los Alamos
National Laboratory, the Max-Planck-Institute for Astronomy (MPIA),
the Max-Planck-Institute for Astrophysics (MPA), New Mexico State
University, University of Pittsburgh, Princeton University, the United
States Naval Observatory, and the University of Washington.
\end{acknowledgements}

\bibliographystyle{aa}
\bibliography{Ha052}

\begin{thebibliography}{14}
\expandafter\ifx\csname natexlab\endcsname\relax\def\natexlab#1{#1}\fi

\bibitem[{{Greenstein}(1975)}]{gd325}
{Greenstein}, J. 1975, \apjl, 196, L117

\bibitem[{{Harris} {et~al.}(2003){Harris}, {Liebert}, {Kleinman}, {Nitta},
  {Anderson}, {Knapp}, {Krzesi{\' n}ski}, {Schmidt}, {Strauss}, {Vanden Berk},
  {Eisenstein}, {Hawley}, {Margon}, {Munn}, {Silvestri}, {Smith}, {Szkody},
  {Collinge}, {Dahn}, {Fan}, {Hall}, {Schneider}, {Brinkmann}, {Burles},
  {Gunn}, {Hennessy}, {Hindsley}, {Ivezi{\' c}}, {Kent}, {Lamb}, {Lupton},
  {Nichol}, {Pier}, {Schlegel}, {SubbaRao}, {Uomoto}, {Yanny}, \&
  {York}}]{wdsloan}
{Harris}, H., {Liebert}, J., {Kleinman}, S., {et~al.} 2003, \aj, 126, 1023

\bibitem[{{Iben} \& {Livio}(1993)}]{cetimescale}
{Iben}, I. \& {Livio}, M. 1993, \pasp, 105, 1373

\bibitem[{{Kalirai} {et~al.}(2004){Kalirai}, {Richer}, {Hansen}, {Reitzel}, \&
  {Rich}}]{kalirai}
{Kalirai}, J., {Richer}, H., {Hansen}, B., {Reitzel}, D., \& {Rich}, R. 2004,
  astro-ph/0409172

\bibitem[{{Kleinman} {et~al.}(2004){Kleinman}, {Harris}, {Eisenstein},
  {Liebert}, {Nitta}, {Krzesi{\' n}ski}, {Munn}, {Dahn}, {Hawley}, {Pier},
  {Schmidt}, {Silvestri}, {Smith}, {Szkody}, {Strauss}, {Knapp}, {Collinge},
  {Mukadam}, {Koester}, {Uomoto}, {Schlegel}, {Anderson}, {Brinkmann}, {Lamb},
  {Schneider}, \& {York}}]{kleinman}
{Kleinman}, S., {Harris}, H., {Eisenstein}, D., {et~al.} 2004, \apj, 607, 426

\bibitem[{{Liebert} {et~al.}(1986){Liebert}, {Wesemael}, {Hansen}, {Fontaine},
  {Shipman}, {Sion}, {Winget}, \& {Green}}]{liebert}
{Liebert}, J., {Wesemael}, F., {Hansen}, C., {et~al.} 1986, \apj, 309, 241

\bibitem[{{Moffett} {et~al.}(1985){Moffett}, {Barnes}, \& {Evans}}]{eg388}
{Moffett}, T., {Barnes}, T., \& {Evans}, D. 1985, in IAU Symp. 111: Calibration
  of Fundamental Stellar Quantities, 365--368

\bibitem[{{Paczynski}(1976)}]{paczynski}
{Paczynski}, B. 1976, in IAU Symp. 73: Structure and Evolution of Close Binary
  Systems, 75--80

\bibitem[{{Pickles}(1998)}]{pickles}
{Pickles}, A. 1998, \pasp, 110, 863

\bibitem[{{Raymond} {et~al.}(2003){Raymond}, {Szkody}, {Hawley}, {Anderson},
  {Brinkmann}, {Covey}, {McGehee}, {Schneider}, {West}, \& {York}}]{raymond}
{Raymond}, S., {Szkody}, P., {Hawley}, S., {et~al.} 2003, \aj, 125, 2621

\bibitem[{{Verbunt} \& {Rappaport}(1988)}]{verbunt}
{Verbunt}, F. \& {Rappaport}, S. 1988, \apj, 332, 193

\bibitem[{{Wagner} {et~al.}(1988){Wagner}, {Sion}, {Liebert}, \&
  {Starrfield}}]{cbs47}
{Wagner}, R., {Sion}, E., {Liebert}, J., \& {Starrfield}, S. 1988, \apj, 328,
  213

\bibitem[{{Wesemael} {et~al.}(1993){Wesemael}, {Greenstein}, {Liebert},
  {Lamontagne}, {Fontaine}, {Bergeron}, \& {Glaspey}}]{wdatlas}
{Wesemael}, F., {Greenstein}, J., {Liebert}, J., {et~al.} 1993, \pasp, 105, 761

\bibitem[{{York} {et~al.}(2000){York}, {Adelman}, {Anderson}, {Anderson},
  {Annis}, {Bahcall}, {Bakken}, {Barkhouser}, {Bastian}, {Berman}, {Boroski},
  {Bracker}, {Briegel}, {Briggs}, {Brinkmann}, {Brunner}, {Burles}, {Carey},
  {Carr}, {Castander}, {Chen}, {Colestock}, \& {Connolly}}]{york}
{York}, D., {Adelman}, J., {Anderson}, J., {et~al.} 2000, \aj, 120, 1579

\end{thebibliography}

\end{document}